\documentstyle[12pt,preprint]{aastex}

\begin{document}

\title{EFFECTS OF A BURST OF FORMATION OF FIRST-GENERATION STARS 
       ON THE EVOLUTION OF GALAXIES}

\author{Yasuhiro Shioya$^1$, Yoshiaki Taniguchi$^1$, Takashi Murayama$^1$, 
        Shingo Nishiura$^2$, Tohru Nagao$^1$, \& Yuko Kakazu$^{3}$}

\affil{$^1$Astronomical Institute, Graduate School of Science,
       Tohoku University, Aramaki, Aoba, Sendai 980-8578, Japan}
\affil{$^2$Kiso Observatory, Institute of Astronomy,
       The University of Tokyo, Mitake-mura, Kiso-gun, Nagano 
       397-0101, Japan}
\affil{$^3$Institute for Astronomy, University of Hawaii,
       2680 Woodlawn Drive, Honolulu, HI 96822}

\begin{abstract}
First-generation (Population III)
stars in the universe play an important role in
early enrichment of heavy elements in galaxies and intergalactic
medium and thus affect the history of galaxies. 
The physical and chemical properties
of primordial gas clouds are significantly different
from those of present-day gas clouds observed in the nearby universe
because the primordial gas clouds do not contain any heavy elements
which are important coolants in the gas.
Previous theoretical considerations have suggested that typical masses
of the first-generation stars are between several $M_\odot$ and 
$\approx 10 M_\odot$ although it has been argued that
the formation of very massive stars (e.g., $> 100 M_\odot$) is also likely.
If stars with several $M_\odot$ are most popular ones
at the epoch of galaxy formation, most stars will evolve to
hot (e.g., $\gtrsim 10^5$ K), luminous ($\sim 10^4 L_\odot$)
stars with gaseous and dusty envelope
prior to going to die as white dwarf stars.
Although the duration of this phase is short (e.g., $\sim 10^5$ 
yr), such evolved stars could contribute both to the ionization
of gas in galaxies and to the production of a lot of dust grains
if the formation of intermediate-mass stars is highly enhanced.
We compare gaseous emission-line properties of such nebulae 
with some interesting high-redshift galaxies such as
IRAS F10214+4724 and powerful radio galaxies.
\end{abstract}


\section {INTRODUCTION}

In the standard big bang cosmology, 
the amount of heavy elements
synthesized in the universe until the decoupling of matter from
primordial radiation is negligibly small; i.e.,
$Z < 10^{-10}$ where $Z$ is the mass fraction of heavy elements
(Wagoner, Fowler, \& Hoyle 1967).
On the other hand, gas clouds in the present-day galaxies have an
appreciable amount of heavy elements 
$Z \sim 0.01$; e.g., the solar heavy element abundance is
$Z_\odot \approx 0.02$. This means that 
chemical and thermal properties of gas clouds in primeval galaxies
at high redshift are significantly different from those in
the present-day galaxies. 

Just after the recombination of the universe, the hydrogen gas
could be in atomic phase at a temperature of $\approx 10^4$ K.
Then the formation of primordial hydrogen molecules can proceed
through the gas phase reaction (e.g., Peebles \& Dicke 1968). 
Since the rotational and vibrational
transitions in hydrogen molecules can cool the gas further,
the first generation stars can be born in such gas clouds
at redshift $\sim 10$ - 30.
Since the temperature is still much higher than a typical kinetic
temperature of molecular gas clouds in the present-day galaxies
(e.g., $\sim 10$ K), the initial mass function (IMF)
in primeval galaxies may be different from that in the present-day
galaxies; i.e., a Salpeter-like IMF (Salpeter 1955; Scalo 1986).
Therefore, one of fundamental questions
on the star formation in primeval galaxies is; ^^ ^^ What kind of stars
(i.e., Population III stars)
were made in such extremely metal deficient gas clouds ?"
[Matsuda, Sato, \& Takeda 1969; Yoneyama 1972; Yoshii \& Saio 1986
(hereafter YS86);
Silk 1977a; Tegmark et al. 1997; Nakamura \& Umemura 1999, 2001a, 2001b;
Bromm et al. 1999, 2001, 2002; Coppi et al. 2001].

Applying the opacity-limited fragmentation theory 
[the fragmentation of gas clouds due to the gravitational instability
continues until the fragments become to be optically thick to 
the cooling radiation (Silk 1977b, 1977c)], YS86 first
derived an IMF for a hydrogen-helium gas cloud.
They found that  typical masses of 
the first-generation stars are from several  $M_\odot$
to 10 $M_\odot$. 
Nakamura \& Umemura (1999) obtained similar results 
based on their one-dimensional hydrodynamical calculations 
together with nonequilibrium processes for hydrogen molecule
formation.
Abia et al. (2001) pointed out that
the very large C and N enhancements observed in the extremely
metal-poor stars in our Galaxy favor a Population III IMF
peaked at intermediate-mass stars.

Recently, on one hand,
some groups investigated the collapse and fragmentation 
of primordial, metal-free gas by numerical simulations 
(Abel, Bryan, \& Norman 2000, 2002; Bromm et al. 1999, 2001, 2002; 
Coppi et al. 2001). 
They showed that the first generation gaseous clumps have 
masses of $\ge 1000M_{\odot}$ 
which corresponds to the Jeans mass of the gas 
with temperature $T \sim $ a few 100 K and the density 
$n \sim 10^3 - 10^4 \; {\rm cm}^{-3}$. 
As Bromm et al. (2002) assumed, if the clump masses are 
indicative of the final stellar mass, Population III 
stars may be more massive than $100 M_{\odot}$. 
Nakamura \& Umemura (2001a, 2001b) have developed 
their hydrodynamical simulations and found that 
the initial mass function of Population III stars 
may be bimodal with peaks of $\approx 1$ -- $10 M_\odot$ and 
$\approx 100 M_\odot$.
The mass depends on the initial density and 
the initial fraction of molecular hydrogen ($x_{\rm H_{2},0}$) 
(Nakamura \& Umemura 2001b). 
If $x_{\rm H_2,0}$ is larger than $\sim 3 \times 10^{-3}$
for low-density ($n \lesssim 10^5$ cm$^{-3}$), 
the effective HD cooling makes the high mass peak small to 
$\approx 10 M_{\odot}$.  However, 
if  $x_{\rm H_2,0}$ is smaller than $\sim 3 \times 10^{-3}$,
very massive stars with $\sim 100 M_\odot$ are formed 
preferentially. 

In this way, the mass of first-generation stars is 
controversial. 
It seems also hard to examine observationally which mode is more feasible;
i.e., Population III stars are either intermediate-mass star or
very massive ones.
If most Population III stars are as massive as $\sim 100 M_\odot$,
they could explode quickly and then contribute to the chemical
enrichment of either galaxies or intergalactic medium or both
(e.g., Madau, Ferrara, \& Rees 2001 and references therein).
Since usual Population II stars could be formed in cold gas
clouds polluted chemically by first Population III stars, 
we cannot examine observationally whether or not
such very massive stars were really formed in the primordial
gas clouds. 
Furthermore, according to Nakamura \& Umemura (2001a), 
proto-galactic clouds with a higher density tend to 
form intermediate-mass stars. If such proto-galactic
systems are present, their evolution can be governed
by evolution of the intermediate-mass stars whose lifetime 
is $\sim 10^8$ yr, being much longer than those of 
very massive stars. 
We investigate what happens and follows in such bursts of
formation of intermediate-mass stars
and discuss effects on the further evolution of galaxies,
adopting the IMF of YS86. 
In particular, intermediate-mass stars experience inevitably
the planetary-nebula-nucleus (PNN) phase, providing a lot of
high-energy photons (e.g., Vassiliadis \& Wood 1994).
 Therefore, it is expected that 
the galaxies appear to be luminous emission-line galaxies
during the PNN phase. We discuss how they look and compare
our model results with observations of some interesting 
high-redshift galaxies.
 
\section{EVOLUTION OF POPULATION-III STARBURSTS IN FORMING GALAXIES}

\subsection{Initial Mass Function for Population III Stars derived by YS86}

In many previous galaxy evolution models,
it has been often assumed that stars are formed with a Salpeter-like 
initial mass function (IMF) in a galaxy;

\begin{equation}
\phi(m) = \beta m^{-\mu}
\end{equation}
where $m$ is the stellar mass in units of $M_\odot$ and
$\beta$ is a normalization constant determined by the relation

\begin{equation}
\int_{m_l}^{m_u} \phi(m) dm = 1,
\end{equation}
which leads to

\begin{equation}
\beta  = \frac{(\mu -1)m_l^{\mu-1}}{1 - (m_l/m_u)^{\mu-1}}.
\end{equation}
In this formulation, 
there are three free parameters; the power index ($\mu$), and the upper 
and lower mass limits of the IMF ($m_u$ and $m_l$).
In order to describe the evolution of the solar neighborhood, 
one adopts conventionally $\mu$ = 1.35, $m_l = 0.1 M_\odot$, 
and $m_u = 60 M_\odot$
or $100 M_\odot$.
However, it is highly uncertain whether or not this 
Salpeter-like IMF is applicable to the star formation history 
in primeval galaxies.

YS86 investigated the IMF for a hydrogen-helium gas 
cloud in detail based on the opacity-limited fragmentation theory. 
They found that  typical masses of the first-generation stars are
from several $M_\odot$ to 10 $M_\odot$, depending on the assumed 
mass-luminosity relation for protostars derived by Silk (1977c);
assuming that the opacity sources within the protostar is due to
grains, he obtained the protostar luminosity

\begin{equation}
L(m) = \epsilon 10^{2.47} (m/M_\odot)^\eta ~ L_\odot,
\end{equation}
where $\epsilon = (Z/Z_\odot)^{-1}$ and $\eta = 3$. 
Since this equation cannot be applicable to  the luminosity of stars
with $Z \lesssim 10^{-4} Z_\odot$, YS86 treated
$\epsilon$ as well as $\eta$ as free parameters in the ranges
$10 \lesssim \epsilon \lesssim 100$ and $1.5 \lesssim \eta \lesssim 3$. 
Then YS86 found that the mass at the peak of the IMF is 
$\approx 4 M_\odot$ for $\eta = 3$ and $\approx 10 M_\odot$ for $\eta = 1.5$.
Another important conclusion in YS86 is that stars more massive than these peaks
are formed less efficiently than those expected from the Salpeter IMF
with $\alpha = 1.35$. In addition, stars less massive than these peaks
are also formed less efficiently than those expected from the same 
Salpeter IMF (see Figure 3 in YS86). Therefore, the IMFs derived by
YS86 are significantly different from the Salpeter IMFs.
We therefore adopt the IMFs of YS86 and investigate luminosity and
spectral evolution of galaxies for various sets of the parameters.

For convenience, we present a parametric form of the YS86 IMF in the
mass range between 1 $M_\odot$ and 60 $M_\odot$ for $\eta = 3$ (YS86a),

\begin{equation}
\phi(m) = \left\{
\begin{array}{lcl}
1.5944 \times 10^{-2} ~ m^{2.0}  & & 1  \le m/M_{\odot} < 2.5 \\
4.1728 \times 10^{-2} ~ m^{0.95} & & 2.5 \le m/M_{\odot} < 3.7 \\
1.4462 \times 10^{-1} ~ m^{0}   & & 3.7 \le m/M_{\odot} < 5  \\
1.0813 ~ m^{-1.25}               & & 5  \le m/M_{\odot} < 7.5 \\
2.0081 \times 10 ~ m^{-2.7}      & & 7.5 \le m/M_{\odot} < 11 \\
2.2088 \times 10^{2} ~ m^{-3.7}  & & 11 \le m/M_{\odot} \le 60 \\
\end{array}
\right.
\end{equation}
and that for $\eta = 1.5$ (YS86b),

\begin{equation}
\phi(m) = \left\{
\begin{array}{lcl}
7.3036 \times 10^{-4} ~ m^{3.5}  & & 1  \le m/M_{\odot} < 2.9 \\
4.4629 \times 10^{-3} ~ m^{1.8}  & & 2.9 \le m/M_{\odot} < 5  \\
8.0865 \times 10^{-2} ~ m^{0}   & & 5  \le m/M_{\odot} < 8  \\
1.2849 ~ m^{-1.33}               & & 8  \le m/M_{\odot} < 12 \\
1.4671 \times 10 ~ m^{-2.31}     & & 12 \le m/M_{\odot} \le 60 \\
\end{array}
\right.
\end{equation}
Here the above numerical coefficients are determined by the 
following normalization for each case, 

\begin{equation}
\int_1^{60} \phi (m) dm = 1.
\end{equation}
In Figure 1, we show the shapes of IMF which we use in this paper. 
Since the massive end slope of YS86a IMF is steepest, 
the number of massive stars ($>20 M_{\odot}$) is smallest 
among the three, although the number of intermediate-mass star 
exceeds that of Salpeter IMF.  

\subsection{How Populous are Intermediate-Mass Stars ?}

Prior going to investigating the luminosity and spectral evolution of 
galaxies using the YS86 IMFs given in equations (5) and (6), we estimate how populous
intermediate-mass stars are in the YS86 IMFs with respect to 
those expected from usual Salpeter-like IMFs.
Since it is expected that the most spectacular effect of
intermediate-mass enhanced star formation is brought by
a number of luminous PNNs with very
high temperature $\sim 10^6$ K (e.g., Vassiliadis \& Wood 1994).
Since the mass of such high-temperature PNN progenitors lies 
in a range between $4 M_\odot$ and $5 M_\odot$, we estimate 
how many stars within this mass range are formed.

If we adopt a Salpeter IMF, the number of stars with a mass range
$m_1 \le m_* \le m_2$ formed in the gas with 1 $M_\odot$ 
is estimated as

\begin{equation}
N (m_1 \le m_* \le m_2)= \int_{m_1}^{m_2} \frac{\phi(m)}{m} dm.
\end{equation}
Using $\beta$ given in equation (3), we re-write equation (8) as

\begin{equation}
N (m_1 \le m_* \le m_2) = \left( \frac{\beta}{\mu} \right)
(m_1^{- \mu} - m_2^{- \mu}) ~~ {\rm stars}~M_\odot^{-1}.
\end{equation}
Taking $m_1 = 4 M_\odot$ and $m_2 = 5 M_\odot$,
we obtain the numbers for the following two cases; 
1) $\mu$ = 1.35, $m_l = 0.1 M_\odot$, and $m_u = 60 M_\odot$, 
and 2) $\mu$ = 1.35, $m_l = 1 M_\odot$, and $m_u = 60 M_\odot$ (see Table 1).

If we adopt a YS86 IMF, we can estimate the number of stars 
with a mass range $4 M_\odot \le m_* \le 5 M_\odot$ by integrating
equations (5) and (6) for $\eta = 1.5$ and $\eta =3$, respectively.
The results are shown in Table 1. We also show the number if 
all stars have a mass of 5 $M_\odot$; i.e., the IMF is
described as a $\delta$ function with $m_* = 5 M_\odot$.

As shown in Table 1, the number of intermediate-mass stars with
$m_* = 4 M_\odot$ -- $5 M_\odot$ for the YS86 IMF with $\eta = 1.5$
is nearly the same as that for the Salpeter IMF with $m_l = 1 M_\odot$.
However, that for the YS86 IMF with $\eta = 3$ is larger by a factor 
of 2.4 than that for the Salpeter IMF with $m_l = 1 M_\odot$.
For both cases, the ratio of the number of intermediate-mass stars with 
$m_*=4M_{\odot}$ -- $5M_{\odot}$ for YS86 IMFs to that for Salpeter 
IMF is within a factor of 3.  
If stars with mass of $5 M_\odot$ are only formed, the number 
is larger by a factor of 15 than that for the Salpeter IMF with 
$m_l = 1 M_\odot$ (see section~3).

\subsection{Galaxy Evolution Models}

In this section, we show the evolution of bolometric luminosity, 
ionizing photon production rate, and emission line ratios of 
Population III star clusters with the YS86 IMFs. 
We adopt the $10^7$ yr burst model as the star formation history. 
We calculate the spectral energy distributions (SEDs) of star clusters 
by using the evolutionary population synthesis code 
PEGASE2 (Fioc \& Rocca-Volmerange 1997, 2000). 
Although the code can be used to calculate the photometric evolution 
of galaxies being the effect of the chemical evolution into account, 
we calculate that of a single metallicity population with $Z=0.0001$; 
note that this metallicity is the poorest metallicity available in PEGASE2. 
The duration of the starburst is longer than the lifetime of the most massive star 
in our model ($60 M_{\odot}$), $5 \times 10^6$ yr (Girardi et al. 1996). 
Since the lifetime is a half of the starburst duration, the metallicity of a 
half of stars formed in the starburst is $Z < 10^{-10}$. 

Figure 2 shows the SED evolution for each model: Salpeter IMF (solid lines), YS86a IMF 
(dotted lines) and YS86b IMF (dashed lines).  
We show the SEDs at 0.01 Gyr, 0.1 Gyr and 0.5 Gyr. 
Note that the ordinate is the flux in units of 
erg s$^{-1}$ $M_{\odot}^{-1}$, corresponding to the case 
in which the total gas mass is $1 M_{\odot}$. 
The SEDs are mainly determined by 
the upper main sequence stars 
and are not sensitive to the IMF.
During the starburst, the ionizing photons are radiated from massive 
main sequence stars whose effective temperature is lower than 
about 60,000 K (Girardi et al. 1996).
Therefore there is a large break at 228 \AA; 
the photons with wavelength shorter than this 
are capable of ionizing the singly ionized helium. 
At an age of $> 10^8$ yr, hottest stars in the star cluster 
are not main sequence stars but post-asymptotic giant branch (AGB) 
stars whose progenitors are stars with a degenerate carbon core, 
e.g., intermediate-mass stars. 
The effective temperature of such post-AGB stars (i.e., PNNs) is about 200,000 K 
(Bl\"ocker 1995). 
Therefore the break of 228 \AA~ of such stars becomes to be small and 
thus the SED shape of far-UV is very hard.  

The global shape of SED depends slightly on the IMF (Figure 3).
During a period of the starburst, between $4 \times 10^6$ yr 
and $\sim 1 \times 10^7$ yr, colors ($U-B$, $B-V$, and $V-K$) 
become redder because of 
red supergiants. 
In this phase, the star cluster with the Salpeter IMF is reddest 
and that with the YS86a is bluest. 
This is because the number ratio of red supergiants to 
upper main sequence stars 
is highest for the Salpeter IMF ($m^{-1.35}$) 
and lowest for the YS86a IMF ($m^{-3.7}$). 
On the other hand, for ages longer than $10^8$ yr, 
the star cluster with the Salpeter IMF is bluest and 
that with the YS86b is reddest. 
The reason for this is that the number ratio of redgiants to 
upper main sequence stars 
is lowest for the Salpeter IMF ($m^{-1.35}$) 
and highest for the YS86b IMF ($m^{3.5}$).

Figure 4 shows the evolution of bolometric luminosity of 
a star cluster of $1 M_{\odot}$ for various IMFs. 
The bolometric luminosity reaches its maximum at the end of 
starburst ($10^7$ yr). 
The star cluster with the YS86b IMF is brightest while that with the YS86a IMF 
is faintest among the models.
After that, following the death of massive stars, 
the bolometric luminosity of stellar systems decreases with time.  
The decreasing rate of the bolometric luminosity also depends 
on the functional form of IMF. 
Since the YS86a IMF has a peak at $m \approx 4 M_{\odot}$ 
whose lifetime is about 0.15 Gyr, 
the decreasing rate of the bolometric luminosity is small till 
the time, and becomes large after that. 
Since the lifetime of stars with $1 M_{\odot}$ is about 5.6 Gyr,  
all the stars formed during the starburst have died till the time.  
As we show below, the dominant sources of the ionizing photons 
become the PNN at the age of $\sim$ 0.1 Gyr. 
The bolometric luminosity to mass ratios decrease 
from $66L_{\odot}/M_{\odot}$ to $3.2L_{\odot}/M_{\odot}$ 
during this period. 

In the upper panel of Figure 5, we show the evolution of production rate of 
photons capable of ionizing neutral hydrogen, $N({\rm H^0})$, 
which is calculated by integrating the spectra below 912 \AA, 
and that of photons capable of ionizing ionized helium, 
$N({\rm He^+})$, which is calculated by integrating the spectra 
below 228 \AA. 
First, we mention how the production rate of the hydrogen ionizing 
photons evolves. 
As the bolometric luminosity increases,
the ionizing photon production rate 
reaches its maximum at the end of starburst, because the lifetime of 
early B stars of $\sim 10M_{\odot}$ is longer than $10^7$ yr. 
The evolution of $N({\rm H^0})$ depends on IMF. 
During the starburst, the ionizing photon production rate per unit mass 
is largest for the star cluster with the YS86b IMF but smallest 
for that with the YS86a IMF.  
In our calculations, we adopt the mass of most massive stars 
 $M_u=60M_{\odot}$ 
whose lifetime is about $5 \times 10^6$ yr. 
When the age of star clusters reaches the lifetime of 
the most massive stars, the number of hydrogen ionizing photons 
begins to decrease with time, following the death of stars 
radiating hydrogen ionizing photons. 
At the age of $\sim 10^8$ yr, the intermediate-mass stars evolve 
to PNNs and they become 
dominant ionization sources for hydrogen. 
In the star cluster with the Salpeter IMF, 
the number of PNNs hardly changes and thus the production rate of 
ionizing photons is kept nearly constant. 
On the other hand, for the star cluster whose IMF is either YS86a or YS86b, 
the number of PNNs and the production rate of ionizing photons 
decrease 
with decreasing number of low mass stars. 
After the age of $\sim$ 1 Gyr, 
the ionizing photon production rate is highest for the star cluster with 
the Salpeter IMF and smallest for that with the YS86b IMF 
since it is proportional to the number of PNNs. 
Next, we comment on the evolution of the production rate of 
ionized helium ionizing photons. 
The lifetime of stars with $m > 10M_{\odot}$ is 
shorter than a few $\times 10^7$ yr. 
After the massive stars have died, 
the remaining main sequence stars cannot ionize 
the neutral helium. 
When the age of the star cluster become $1 \times 10^8$ yr, 
PNNs come into existence. 
The effective temperature of PNNs is very high enough to radiate 
He$^+$ ionizing photons. 

The lower panel of Figure~5 shows the evolution of the ratio of 
$N({\rm He^+})$ to $N({\rm H^0})$ which indicates the hardness of 
UV spectra. 
During the starburst, the ratio is lower than $10^{-3}$. 
However, after PNNs become the dominant source of UV radiation, 
the ratio exceeds $10^{-3}$. 
The stellar mass of a progenitor of PNN decreases with time. 
Therefore, the effective temperature of PNN decreases and thus 
the ratio of $N({\rm He^+})$ to $N({\rm H^0})$ also decreases 
(Vassiliadis \& Wood 1994). 

We investigate ultraviolet emission-line properties 
of the gaseous nebula photoionized by PNN clusters during the 
course of galaxy evolution.
We use the photoionization code CLOUDY94 (Ferland 1997), 
which solves the equations of statistical and thermal equilibrium 
and produces a self-consistent model of the run of temperature 
as a function of depth into the nebula. 
Here we assume that a uniform density, dust-free gas cloud with 
plane-parallel geometry is ionized by various continuum sources. 
The parameters for the calculations are (1) the hydrogen density 
of the cloud, $n_{\rm H}$, (2) the ionization parameter, 
$U=Q({\rm H^0})(4 \pi r^2 n_{\rm H}c)^{-1}$, where $Q({\rm H^0})$ 
is the number of ionizing photons, $r$ is the distance from 
the ionizing source, and $c$ is the speed of light (Osterbrock 1989), 
(3) the SED of the ionizing radiation, and (4) the chemical 
composition. 
The SED of the ionizing radiation is derived as noted before 
using PEGASE2 (see Figure 2). 
We perform our photoionization calculations as a function 
of the ionization parameter between $\log U=-4$ and $-1$ 
with a logarithmic interval of 0.5. 
We adopt $n_{\rm H}=10^2 \; {\rm cm}^{-3}$. 
We also adopt a metallicity of $Z = 0.001 = 0.05 Z_{\odot}$. 
The calculations were stopped when the temperature fell to 3000 K, 
below which little optical emission is expected. 

For reference, we also calculate AGN photoionization models
and shock-heating models in the following ways.
a) AGN photoionization models are calculated by using 
CLOUDY94. We adopt that 
the input SED has a power-law form with $\alpha = -1.5$ 
($f_{\nu} \propto \nu^{\alpha}$).
b) Shock-heating models are taken from results 
of Dopita \& Sutherland (1996). 
Their models assume solar metallicities, 
shock velocities from 
$150 \; {\rm km \; s^{-1}}$ to $500 \; {\rm km \; s^{-1}}$, 
and the magnetic parameter 
$0 \le B/\sqrt{n} \le 4 \mu{\rm G \; cm^{3/2}}$. 
The PNN photoionization model results are shown 
together with these models in Figure 6; 
the UV emission-line diagnostic diagrams 
proposed by De Breuck et al. (2000). 

Since high-energy photons supplied from the PNN clusters 
can cause high-ionization lines such as C {\sc iv},
the UV emission-line properties of the PNN photoionization models are quite similar to 
those of the AGN photoionization. 
In particular, the model locus predicted for $10^7$ yr-burst
stellar populations with age of 0.5 Gyr cannot be distinguished
from that for the AGN models with a power-law continuum with $\alpha = -1.5$.
This similarity has been already utilized to show that 
some LINERs (low-ionization nuclear emission-line regions:
Heckman 1980) can be explained by photoionization by 
PNN clusters (Taniguchi, Shioya, \& Murayama 2000).
On the other hand,
the shock-heating models appear to show different 
properties from those of the PNN and AGN models.

Figure 7 shows the $L_{\rm HeII}/L_{\rm Bol}$, 
Ly$\alpha$/He {\sc ii}, and N {\sc v}/He {\sc ii} 
ratios as a function of $\log U$ for PNN models with age of 0.1 Gyr (open circles) 
and 0.5 Gyr (filled circles). 
These ratios hardly depend on the IMF but depend on the age. 

Recently, the evolutionary tracks of zero-metal stars have been 
published (e.g., Marigo et al. 2001). 
Based on these new models, Schaerer (2002) demonstrated that
massive population III stars can radiate the He$^+$ ionizing photons
because of both 
the high effective temperature ($T_{\rm eff} > 80,000$ K) of 
zero-metal stars and the effect of the non-LTE stellar atmosphere 
(see also Tumlinson \& Shull 2000).
The small $N({\rm He^+})/N({\rm H^0})$ ratio during the starburst 
shown in Fig.5 is attributed to the use of 
the population synthesis model with metallicity of $Z=0.0001$ 
instead that with  $Z=0$.   
We note, however, H {\sc ii}  regions around massive population III stars 
cannot radiate metal lines in which we are interested
since the gas around massive Population III stars is still primordial.

\subsection{Comparison with high-z star forming galaxies}

If the star formation in some forming galaxies takes place with YS86 IMF, 
such galaxies could experience a PNN-dominated phase $\sim 10^8$ yr after 
the initial starburst. 
Therefore it seems important to look for such galaxies at high redshift. 
As is shown in section 2.3, the SED of star clusters in PNN-dominated phase 
is so hard that the emission-line properties 
of the gaseous nebula photoionized by PNN clusters are indistinguishable from 
those photoionized by AGN. 
Considering the stellar evolution of intermediate mass stars, 
the PNN phase follows the post-AGB phase when stars eject a dusty gaseous envelope. 
It is therefore expected that the PNN cluster is surrounded by dusty gas 
(see the next section). 
Until the PNN-dominated phase begin, dust formed in the ejecta of 
Type II supernovae is also present in the interstellar medium 
(Todini \& Ferrara 2001). 
Therefore, a part of ionizing photons could be absorbed by such 
dust grains. However, we neglect this effect in the following discussion
for simplicity.

The above expected features of the 
PNN star cluster are very interesting because 
the PNN star cluster may be observed as a dusty, ultraluminous
infrared galaxy with high-ionization emission lines at high redshift. 
It is known that some high-$z$ forming galaxy candidates share nearly the same
properties; e.g., IRAS F 10214$+$4724 at $z = 2.3$ 
(Rowan-Robinson et al. 1991) and
SMM 02399$-$0136 at $z = 2.8$ (Ivison et al. 1998).
Their observed huge luminosities  ($L_{\rm bol} \gtrsim 10^{12}
L_\odot$) have been often considered to be powered by massive stars.
However, high-ionization emission lines such as C {\sc iv} $\lambda$1549
\AA ~ and N {\sc v} $\lambda$1240 \AA ~ are also  observed
in them. Since these emission lines cannot be produced by photoionization
by ordinary massive stars, it has been often considered that 
active galactic nuclei (AGNs) also reside 
in them (Ivison et al. 1998; Serjeant et al. 1998).
In particular, the presence of hidden broad 
emission lines reinforces this interpretation for 
IRAS F10214$+$4724 (Goodrich et al. 1996). 
Spectropolarimetry of SMM J02399$-$0136 reveals that 
the polarization behavior is similar to BAL quasars (Vernet \& Cimatti 2001). 
Even in the local universe, bursts of star formation
(starbursts) are sometimes associated with galaxies hosting AGNs and thus
the simultaneous presence of both starburst and AGNs may not
be unusual (e.g., Taniguchi 1999; Mouri \& Taniguchi 2001;
Storchi-Bergmann et al. 2001).
However, the evidence for AGNs in them
has been obtained often  by rest-frame ultraviolet and optical
emission lines which are subject to misunderstanding in some cases
(Taniguchi et al. 1999; Lutz, Veilluex, \& Genzel 1999).
Since the effective temperatures of PNNs exceed
$10^5$ K, high energy photons 
are much more numerous than those from ordinary
massive stars. Here we examine whether or not the PNN cluster
is responsible for the formation of high-ionization emission lines.

In Figure 6, we plot the observed positions of IRAS F10214+4714 
(Serjeant et al. 1998) and 
the high-$z$ powerful radio galaxies studied by De Breuck et al. (2000). 
Although the high-$z$ powerful radio galaxies have an AGN, 
at least some of them are considered to be very young galaxies 
(e.g., Eales et al. 1993).
We also plot  available data of planetary nebulae (Feibelman 2000). 
It is found that the emission line ratios of IRAS F10214+4724 
as well as the high-$z$ powerful radio galaxies 
in these UV diagnostics can be explained either by 
the AGN models 
with $-2 \le \log U \le -1.5$
or by the PNN model with an age of 0.5 Gyr. 
The similar behaviors between the PNN and AGN photoionization models
unable us to distinguish the two models unfortunately.

It is noted that both the AGN and PNN models appear to underpredict
the C {\sc ii}]/C {\sc iii} ratio as shown in the right-lower
panel in Figure 6. This discrepancy 
suggests that the observed UV emission-line ratios of 
IRAS F10214+4714 and the  high-$z$ powerful radio galaxies
can be explained either by a combination 
between the shock heating and the AGN  photoionization
or by a combination between  the shock heating and the
PNN photoionization.

Next, we study whether or not the star clusters 
dominated by PNNs can explain 
the luminosity of high-$z$ forming galaxy candidates. 
As shown in section 2.2 the bolometric luminosity to mass ratio 
changes from $66 L_{\odot}/M_{\odot}$ to $3.2 L_{\odot}/M_{\odot}$. 
Therefore, the mass of stellar system needed 
to reproduce the bolometric luminosity of IRAS F10214+4724 
($\sim 10^{13}L_{\odot}$), is $10^{11}-10^{12}M_{\odot}$. 
This value is comparable to the stellar mass of massive galaxies. 
It is impossible to radiate the huge luminosity from only 
PNNs of first generation stars for this galaxy.
There are two reasons why it is hard to explain the huge luminosity
by a lot of PNNs. 
As shown in section~2, even for the YS86's IMF 
the number of intermediate-mass stars (e.g., $4M_{\odot} < M_* <5M_{\odot}$)
is only a few time larger than that for the Salpeter IMF.
The other reason is related to the difference of the lifetime of 
progenitor stars.  
The lifetime of stars with mass of $4M_{\odot}$ is 
$1.54 \times 10^8$ yr and that of $5M_{\odot}$ is $9.9 \times 10^7$ yr 
(Girardi et al. 1996). 
Therefore, the mass range of stars which evolve to PNNs  
at the same time is very small and as a result 
the number of those stars is very small.  
To solve this discrepancy, we consider an extreme model in the next section. 

\section{AN EXTREME MODEL}

\subsection{Scenario}

Here, we consider an extreme case that the formation of intermediate-mass 
stars occurs during a period of $\tau_{\rm SF}$. 
Most luminous forming galaxy candidates at high redshift 
contain a lot of molecular gas (e.g., $M_{\rm gas} \sim
10^{11} M_\odot$) and their inferred star formation rates (SFR)
are of the order of $\sim 10^3 M_\odot$ yr$^{-1}$ 
(Frayer et al. 1998, 1999).
Therefore, we assume that a galaxy has 
a gas mass of $10^{11} M_\odot$ and the initial star formation
occurs at a rate of $SFR \equiv \dot{M}_{\rm SF} = 10^3 M_\odot$ yr$^{-1}$. 
We adopt a duration of the star formation 
$\tau_{\rm SF} = 10^7$ yr. The gas mass  
converted to stars is estimated as $M_{\rm SF} =
\dot{M}_{\rm SF} \times \tau_{\rm SF} = 10^{10} M_\odot$.  
Thus in this model, the star formation efficiency is $\eta_{\rm SF} =$ 10\%.
The star formation may take place in a central region
(say, within a radius of $\sim$ 1 kpc) of galaxies.
We assume for simplicity that all stars formed in this
initial star formation have a mass of 5 $M_\odot$
(i.e., a typical intermediate-mass star).
Therefore, the SFR adopted above means that 
200 stars with $m_* = 5 M_\odot$
are formed in a year; $\dot{N}_* = \dot{M}_{\rm SF}/m_* = 200$
stars yr$^{-1}$.

Here we consider the evolution of a cluster of such stars. 
The bolometric luminosity of each star in the main sequence
phase is $L_* \approx 1 \times 10^3 L_\odot$ and the age of the star
is $\tau_{\rm age} \approx 1 \times 10^8$ years. 
The total number of stars formed in the duration is
$N_*$(total) = $\dot{N}_* \times \tau_{\rm SF} 
= 2 \times 10^9$ stars. Therefore, the total luminosity 
of the star cluster increases during the star formation 
period ($10^7$ years) at a rate of $\dot{N}_* \times L_*
\approx  2 \times 10^5 L_\odot$ yr$^{-1}$ and reach the maximum
bolometric luminosity 
$L_{\rm bol}({\rm max}) \approx  2 \times 10^{12} L_\odot$ 
at $t = 10^7$ years. This maximum luminosity keeps
as long as the stars are alive; i.e., up to $t = \tau_{\rm age} =
10^8$ years. Shortly after this age, the stars evolve from the main 
sequence to PNNs through AGB. 

Since one PNN has a bolometric luminosity
of $L_{\rm PNN} \approx  10^4 L_\odot$ (e.g., Vassiliadis \& Wood 1994),
the total PNN luminosity amounts
to $L_{\rm PNN}({\rm total}) = \dot{N}_* ~ \tau_{\rm PNN} ~ L_{\rm PNN} =
2 \times 10^{11} L_\odot$ where $\tau_{\rm PNN} \sim 10^5$ yr is the 
lifetime of the PNN phase. Nevertheless this short lifetime,
note that the PNN phase
in the star cluster continues during nearly the same period as that of
star formation; i.e., $10^7$ yr.
In the PNN phase, the total luminosity of the star cluster is
mostly dominated by unevolved main sequence stars while the 
ionizing photons are supplied mainly by PNNs.

Another important property of the PNN phase is that a lot of gas
is ejected from the stars. The core mass of each star is 
estimated to be $m_{\rm c} \approx 0.6 M_\odot$. Therefore, each star
ejects a metal-enriched gas of 
$m_{\rm gas, PN} = m_* - m_{\rm c} \simeq 5 M_\odot
- 0.6 M_\odot \simeq 4.4 M_\odot$.
The total ejected gas mass amounts to $M_{\rm gas, PN} =
m_{\rm gas, PN} N_*({\rm total}) \simeq 8.8 \times 10^9 M_\odot$.
Since the gas-to-dust mass ratio in such gas may be $\sim$ 100,
the star cluster can produce a lot of dust grains with a 
mass of $M_{\rm dust} \sim 10^8 M_\odot$.

In summary, the PNN phase of the star cluster has the following
properties; 1) the bolometric luminosity is 
$L_{\rm bol} \sim 10^{12} L_\odot$;
2) the ionizing luminosity is $\sim 2 \times 10^{11} L_\odot$; 
3) the ionizing continuum is so hard that it creates highly-ionized
regions around the star cluster; 
4) the dust mass is $M_{\rm dust} \sim 10^8 M_\odot$; 5) the heavy element
abundance is increased up to $Z \sim M_{\rm dust}/M_{\rm gas}
\sim 0.001$,  and 6) this phase lasts
for $\approx 10^7$ yr  given that the duration of 
star formation is $10^7$ yr.

\subsection{Emission-line properties of the PNN cluster}

As shown in section 2.3, the PNN cluster is capable of 
the formation of the high-ionization lines such as C {\sc iv}
$\lambda$1549 and C {\sc iii}] $\lambda$1909 emission lines.
For example, the observed  C {\sc iv}/He {\sc ii} ratios are
5.7 and 1.9 for SMM 02399$-$0136 and IRAS F10214$+$4724, 
respectively (Ivison et al. 1998; Serjeant et al. 1998).
However, our models cannot explain the  observed higher ratios of
N {\sc v}$\lambda$1240/He {\sc ii}; 8.5 for  SMM 02399$-$0136
(Ivison et al. 1998) and 1.4 for IRAS F10214$+$4724 (Serjeant et al. 1998) (see Figure 7).
In our model, we have assumed that 
each  heavy element abundance is five percent of its solar value.
If only the helium burning occurs in the core of PNN, carbon 
could be enriched. However, if the hydrogen burning also occurs,
nitrogen could be enriched because nitrogen is formed as the end
product of CNO cycle in the core. If such nitrogen enrichment occurs, 
the observed strong N {\sc v} emission may be explained.
It is worthwhile noting that the overabundance of nitrogen is also
suggested for the broad emission-line gas in high-$z$ quasars
(Hamann \& Ferland 1993, 1999).

\subsection{Further Evolution of the Star Cluster}

We consider the final fate of the star cluster in our model
given in section 3.1.
The intermediate-mass stars formed in the initial starburst
will evolve to white dwarf (WD) stars with a mass of $m_{\rm WD}
= m_{\rm c} \approx 0.6 M_\odot$ through the phases of AGB
and planetary nebula. Since the evolved WD stars have a 
luminosity of $\sim 10^{-4} L_\odot$ at their ages of 
$\approx 10^{10}$ yr, the star cluster
formed in the starburst becomes finally very faint; i.e,
$L_{\rm WD}({\rm total}) =  L_{\rm WD} N_*({\rm total})
\simeq 2 \times 10^5
L_\odot$. Since there is no such candidate in the local universe,
it is expected that further star formation could have occurred in
galaxies even if starbursts with only intermediate-mass
stars occurred in the galaxies. After the death of PNN stars,
dust grains with a mass of $\sim 10^8 M_\odot$ are left in
a galaxy. Since the galaxy has the gas with a mass of $10^{11} 
M_\odot$, the heavy element abundance increases at a level of
$Z \sim 0.001$ after the PNN phase.  Therefore, so-called 
Population II stars would form in the gas clouds 
polluted chemically by the intermediate-mass stars. In other word, 
galaxies with a PNN cluster may evolve to ordinary galaxies
observed in the present-day universe. 

\subsection{Final Remarks}

What we would like to propose in this section is not that
some particular high-$z$ dusty galaxies are indeed the PNN
galaxies but that  candidates of the {\it extreme} PNN galaxies share
nearly the same properties of the observed high-$z$ dusty galaxies.
Recent deep surveys both at FIR and submillimeter wavelengths
have discovered 
a significant number of hyperluminous dusty galaxies
at high redshift (Rowan-Robinson et al. 1991; Kawara et al. 1998;
Puget et al. 1999; Smail, Ivison, \& Blain 1997; Hughes et al. 1998;
Barger et al. 1998; Eales et al. 1999; Rowan-Robinson 2000).
If some proto-galactic systems tend to form intermediate-mass
stars preferentially as Population III stars,
it is possible that candidates
of the PNN galaxies may be found in some high-$z$
dusty galaxies. 
If present, such galaxies will be detected in future sensitive search for
dusty galaxies made by Atacama Large Millimeter Array (ALMA;
Kurz \& Shaver 1999). Deep optical spectroscopy using new
30-m class telescopes will be also necessary to investigate
detailed rest-frame UV spectra of such high-$z$ dusty galaxies.

\vspace{0.5cm}

We would like to thank Gary Ferland for making his code 
CLOUDY available to the public. 
We would like to thank Hideyuki Saio and Satoru Ikeuchi 
for useful discussion. We would also like to thank an 
anonymous referee for his/her useful comments and suggestions.
This work was financially supported in part by Grant-in-Aids for the Scientific
Research (Nos. 10044052, and 10304013) of the Japanese Ministry of
Education, Culture, Sports, and Science.
YS is JSPS fellow.


\newpage

\begin{deluxetable}{lc}
\tablenum{1}
\tablecaption{%
The number of intermediate-mass stars with 
$4 M_\odot \leq m_* \leq 5 M_\odot$.
}
\tablehead{
  \colhead{IMF} &
  \colhead{$N(4 M_\odot \leq m_* \leq 5 M_\odot)$}
}
\startdata
Salpeter ($M_l=0.1M_{\odot}$) & 0.0052 \nl
Salpeter ($M_l=1 M_{\odot}$)  & 0.0136 \nl
YS86 ($\eta=3$)(YS86a)        & 0.0323 \nl
YS86 ($\eta=1.5$)(YS86b)      & 0.0149 \nl
$\delta$-function ($5M_{\odot}$) & 0.2 \nl
\enddata
\end{deluxetable}

\newpage

\begin{figure}
\epsscale{1.00}
\plotone{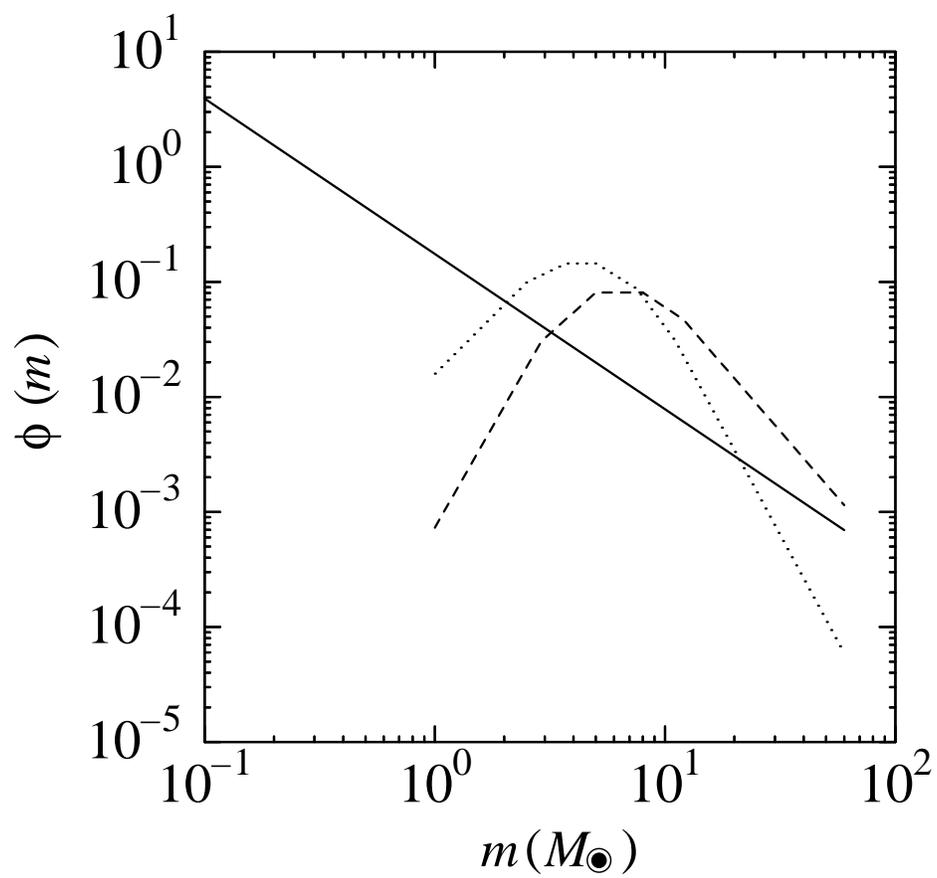}
\caption{Comparisons of YS86 IMFs (dotted line: YS86a, 
dashed line: YS86b) with the Salpeter IMF (solid line).
\label{fig1}}
\end{figure}

\begin{figure}
\epsscale{1.00}
\plotone{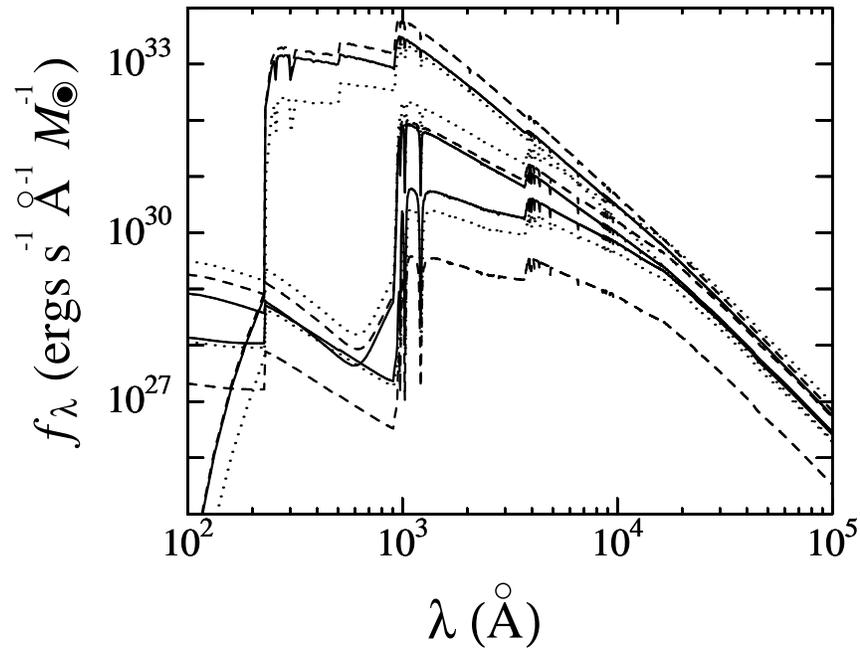}
\caption{The spectral energy distributions 
of star clusters with age of 0.01 Gyr (upper lines), 
0.1 Gyr (middle lines) and 0.5 Gyr (lower lines) for 
Salpeter IMF (solid lines), YS86a IMF 
(dotted lines) and YS86b IMF (dashed lines). 
\label{fig2}}
\end{figure}

\begin{figure}
\epsscale{1.00}
\plotone{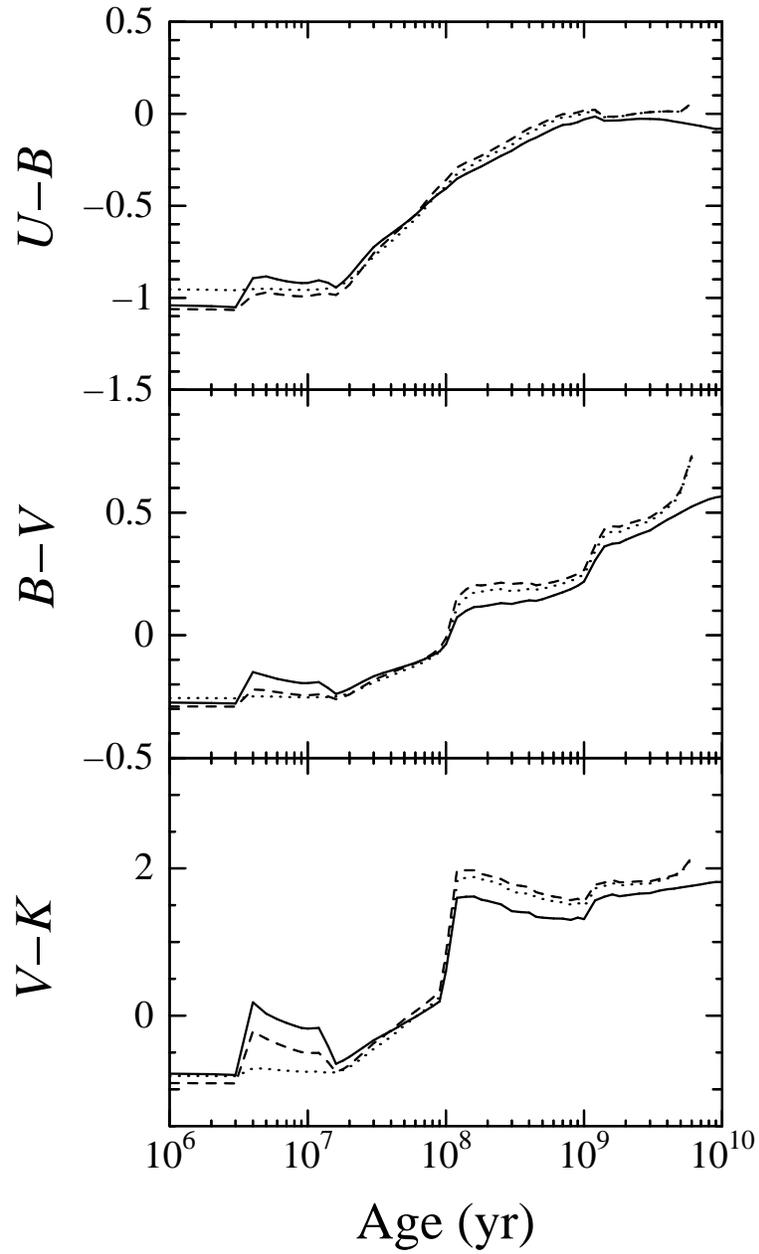}
\caption{Evolution of $U-B$, $B-V$ and $V-K$ colors of 
$10^7$ yr starburst model with Z=0.0001 for various IMFs: 
Salpeter (solid lines), YS86a (dotted lines) and 
YS86b (dashed lines).
\label{fig3}}
\end{figure}

\begin{figure}
\epsscale{1.00}
\plotone{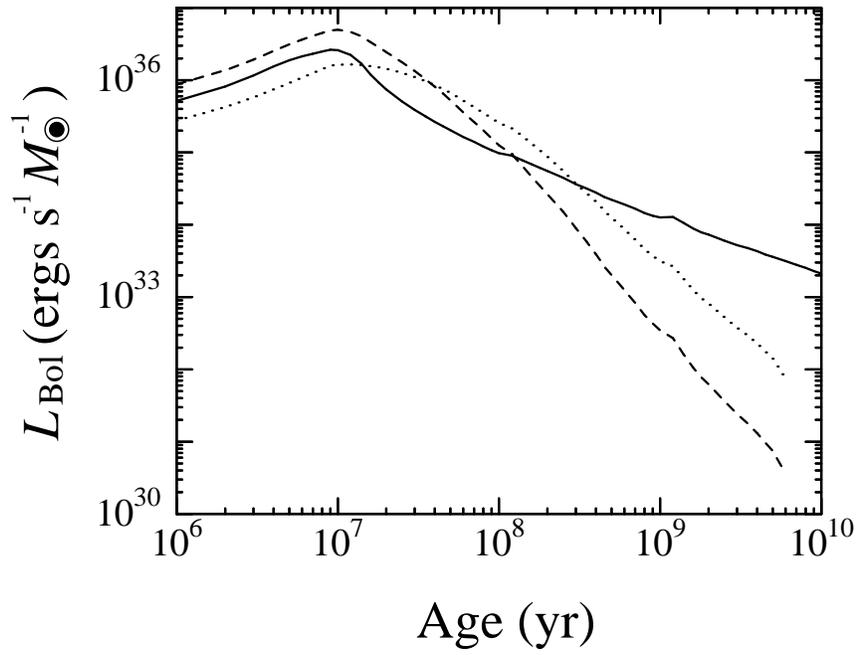}
\caption{Evolution of bolometric luminosity of star clusters 
of mass of $1M_{\odot}$ for Salpeter IMF (solid line), 
YS86a IMF (dotted line) and YS86b IMF (dashed line).
\label{fig4}}
\end{figure}

\begin{figure}
\epsscale{0.70}
\plotone{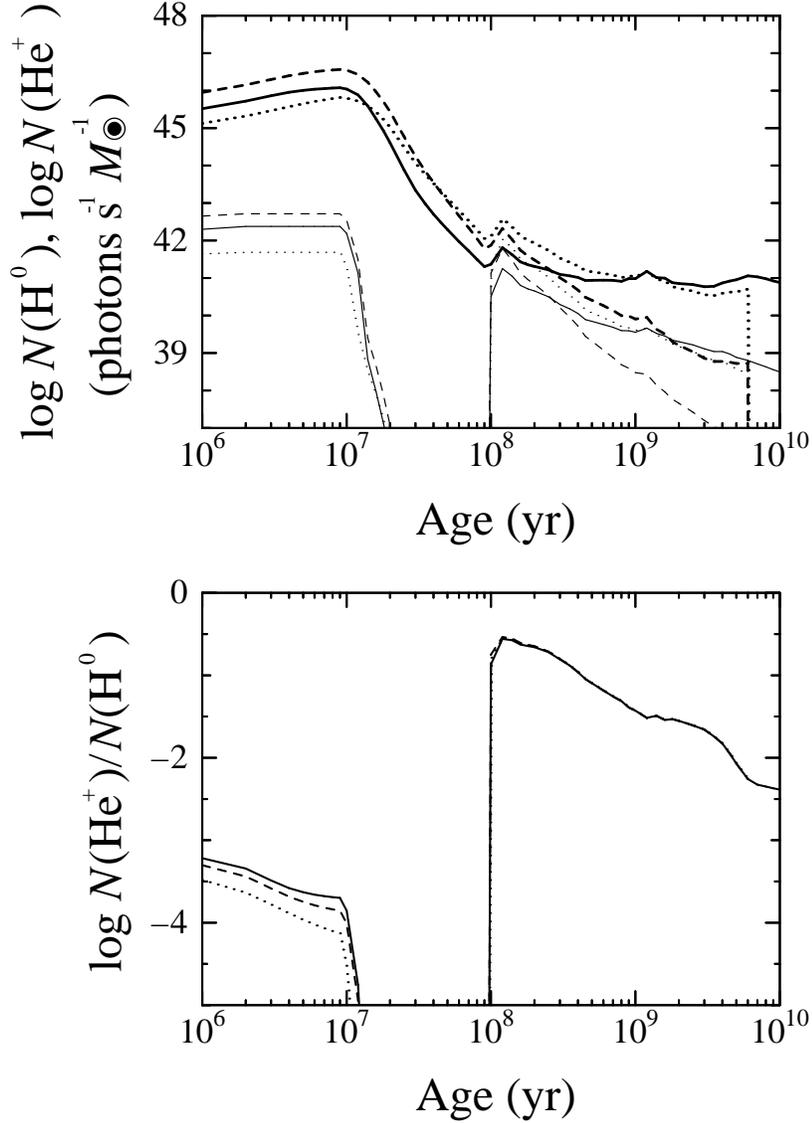}
\caption{(upper) Evolution of ionizing photon production rate 
with wavelength below 912 \AA ($N({\rm H^0})$, thick lines) and that with 
wavelength below 228 \AA ($N({\rm He^+})$, thin lines) for various IMFs: 
Salpeter (solid lines), YS86a (dotted lines) and 
YS86b (dashed lines).
(lower) Evolution of the ratio of $N({\rm He^+})$ to $N({\rm H^0})$ 
for various IMFs: Salpeter (solid lines), YS86a (dotted lines) and 
YS86b (dashed lines).
\label{fig5}}
\end{figure}

\begin{figure}
\epsscale{0.70}
\plotone{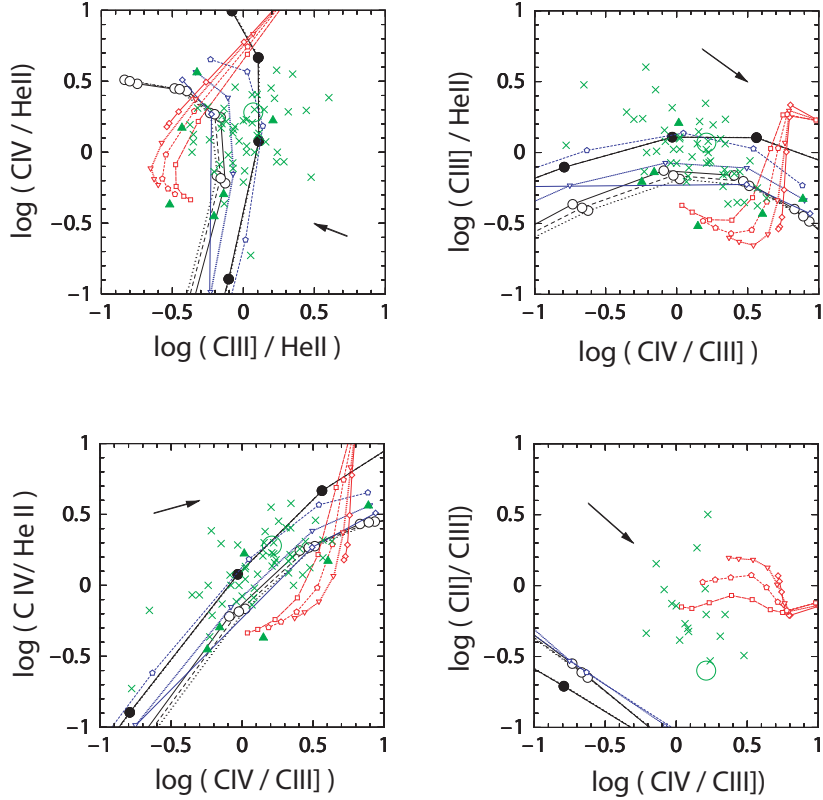}
\caption{Ultraviolet emission line diagnostics adopted 
in De Breuck et al. (2000) to discuss an excitation mechanism. 
Black lines with black marks are PNN models at 0.1 Gyr (open circles) 
and 0.5 Gyr (filled circles) for various IMF: 
Salpeter (solid lines), YS86a (dotted lines) and 
YS86b (dashed lines). 
Blue lines with blue marks are AGN models, power law with 
$\alpha = -1.5$ ($f_{\nu} \propto \nu^{\alpha}$) for various density: 
$n_e=10^2 \; {\rm cm}^{-3}$ (solid lines), 
$n_e=10^3 \; {\rm cm}^{-3}$ (dotted lines), and 
$n_e=10^4 \; {\rm cm}^{-3}$ (dashed lines). 
Red lines with red marks are shock model calculated 
by Dopita \& Sutherland (1996) for various magnetic parameter: 
$B/\sqrt{n}=0$ (solid line), 1 (dotted line), 2 (short dashed line) 
and 4 (long dashed line) $\mu$ G cm$^{3/2}$.    
Green marks show observational points: 
IRAS F10214+4724 (circle), high-z radio galaxies (crosses, De Breuk 
et al. 2000), and planetary nebulae (filled triangle, Feibelman 2000).
The dereddening vector ($A_V=1$) is shown in each panel.
\label{fig6}}
\end{figure}

\begin{figure}
\epsscale{0.70}
\plotone{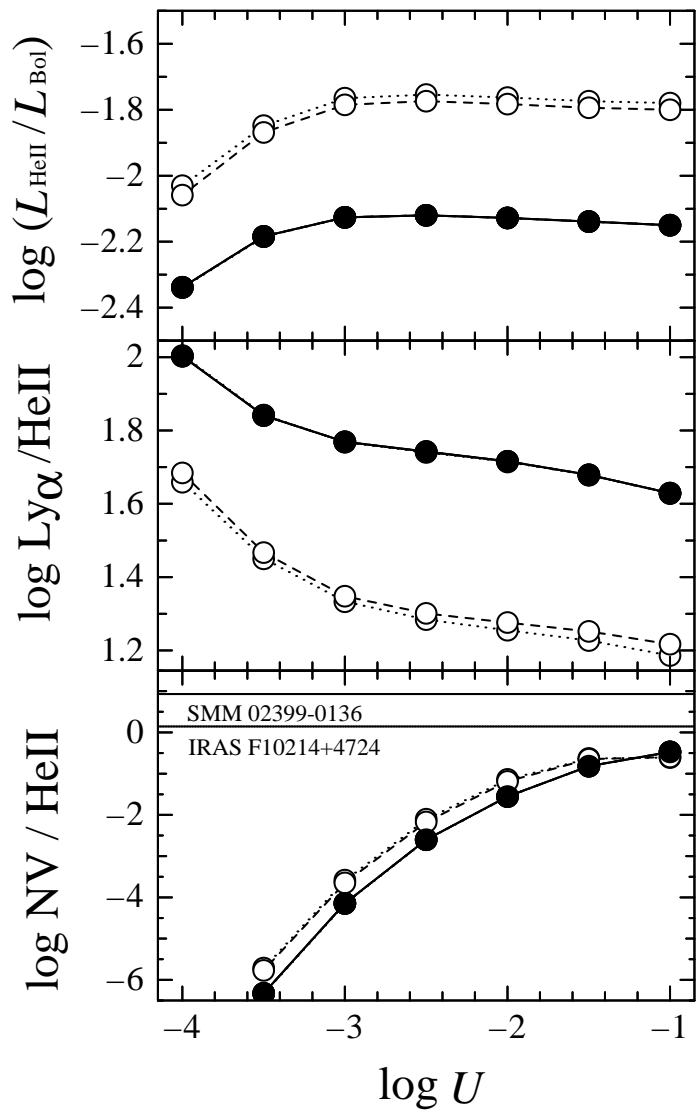}
\caption{Dependence of $\log (L_{\rm HeII}/L_{\rm Bol})$, 
$\rm \log Ly \alpha/He {\sc ii}$, and $\rm \log N {\sc v}/He {\sc ii}$ on 
$\log U$ in our PNN models at 0.1 Gyr (open circles) 
and 0.5 Gyr (filled circles) for various IMF: 
Salpeter (solid lines), YS86a (dotted lines) and 
YS86b (dashed lines)
\label{fig7}}
\end{figure}


\end{document}